\documentclass[%
 reprint,
superscriptaddress,
 amsmath,amssymb,
 aps,
]{revtex4-2}

\usepackage{graphicx}
\usepackage{dcolumn}
\usepackage{bm}
\usepackage{hyperref}
\hypersetup{hypertex=true,
	colorlinks=true,
	linkcolor=blue,
	anchorcolor=blue,
	urlcolor=blue,
	citecolor=blue}
\usepackage{amsmath}

\begin{document}

\preprint{APS/123-QED}

\title{Tuning Layer Orbital Hall Effect via Spin Rotation in Ferromagnetic Transition Metal Dichalcogenides}

\author{Shilei Ji}
\email{jishilei@njupt.edu.cn}
\affiliation{Nanjing University of Posts and Telecommunications (NJUPT), Nanjing 210023, China.}

\author{Jianping Yang}
\affiliation{Nanjing University of Posts and Telecommunications (NJUPT), Nanjing 210023, China.}

\author{Li Gao}
\affiliation{Nanjing University of Posts and Telecommunications (NJUPT), Nanjing 210023, China.}

\author{Xing'ao Li}
\email{lxahbmy@126.com}
\affiliation{Nanjing University of Posts and Telecommunications (NJUPT), Nanjing 210023, China.}
\affiliation{Zhejiang University of Science and Technology, Hangzhou 310023, China.}


\date{\today}

\begin{abstract}
Orbitronics, which leverages the angular momentum of atomic orbitals for information transmission, provides a novel strategy to overcome the limitations of electronic devices. Unlike electron spin, orbital angular momentum (OAM) is strongly influenced by crystal field effects and band topology, making its orientation difficult to manipulate with external fields. In this work, by using first principle calculations, we investigate quantum anomalous Hall insulators (QAHIs) as a model system to study the layer orbital Hall effect (OHE). Due to band inversion, only one valley remains orbital polarization, and thus the OHE originates from a single valley. Based on stacking symmetry analysis, we investigated both AA and AB stacking configurations, which possess mirror and inversion symmetries, respectively. The excitation of OAM exhibits valley selectivity, determined jointly by valley polarization and orbital polarization. In AA stacking, the absence of inversion center gives rise to intrinsic orbital polarization, leading to OAM excitations from different valleys in the two layers. In contrast, AB stacking is protected by inversion symmetry, which enforces valley polarization and causes OAM in both layers to originate from the same valley. Furthermore, the direction of spin polarization tunes the sign of the Berry curvature, thereby dictating the transport of OAM. As a result, in bilayer antiferromagnetic QAHI systems, orbital currents display a distinct layer-contrasting behavior in both flow direction and OAM accumulation.

\end{abstract}

\maketitle


\section{Introduction}\label{secI}

In solids, the motion of electrons is strongly influenced by the crystal field created by surrounding atoms, which restricts the rotational freedom of atomic orbitals.\cite{RN776, RN746, RN778, RN1205, RN1206} This restriction breaks the original orbital degeneracy and leads to orbital quenching, thereby suppressing the formation of net orbital magnetic moments. Nevertheless, in many non-centrosymmetric crystals such as two-dimensional (2D) transition metal dichalcogenides (TMDCs) MoS$_{2}$ and VSe$_{2}$, the breaking of inversion symmetry allows non-zero OAM to emerge locally in momentum space, even though the total orbital magnetization across the Brillouin zone remains zero due to time-reversal symmetry.\cite{RN734, RN733, RN946, RN738, RN986} The crystal field splitting near the valleys leads to pronounced orbital hybridization, in which the $d_{xy}$ and $d_{x^2-y^2}$ orbitals become strongly mixed.\cite{RN1068, RN946, RN1084, RN944} These orbitals carry magnetic quantum numbers $m_l = \pm 2$. Since Berry curvature and OAM obey the same symmetry constraints, regions with finite OAM also exhibit non-zero Berry curvature.\cite{RN734, RN733, RN1125, RN730} The linear combination of atomic orbitals constructs Bloch states, and the resulting electronic wavefunctions retain both spin and orbital features. When driven by an external electric field or a temperature gradient, Bloch electrons with orbital character acquire an anomalous velocity under the influence of Berry curvature. This deflection generates transverse orbital currents, giving rise to the OHE\cite{RN1211,RN1097,RN1093}, orbital Rashba Effect\cite{RN952, RN1208,RN781},  orbital Nernst effect\cite{RN1125,RN1009,RN1107}, which provide novel avenues for controlling orbital degrees of freedom in crystalline solids and exploring their role in transport phenomena.

The band topology of TMDCs has a profound influence on the OHE at their valleys.\cite{RN1068, RN804, RN749, RN946, RN741, RN1202, RN1006, RN1117, RN1033, RN944, RN947} In 2D higher-order topological insulators, which host zero-dimensional corner states, the bulk can sustain a giant orbital Hall conductivity (OHC).\cite{RN946, RN741, RN1033} This effect arises from the opposite OAM and Berry curvature contributions of the two valleys, resulting in the accumulation of OAM with opposite signs at the opposite ends of the sample, reminiscent of the spin Hall effect.\cite{RN734, RN733, RN1084, RN947} However, the robust transport properties of such higher-order topological insulators are hard to manipulate, which limits their applicability as computational devices.\cite{RN946, RN741, RN948, RN1033, RN1084} When the band gap narrows and a band inversion occurs at one of the valleys, the system undergoes a transition to a QAHI.\cite{RN557, RN944, RN792, RN642, RN686, RN1098, RN960} In this phase, the orbital hybridization of the valley shifts into the conduction band, leading to the disappearance of its OAM contribution, while the other valley continues to carry finite OAM. As a result, the orbital Hall response evolves from a two-valley contribution to a single-valley dominated effect, making the OHC directly determined by the valley ordering of the system.\cite{RN946, RN944}

In this work, we investigate 2D QAHIs as a platform to explore the tuning and transport of OAM. By combining first principle calculations with analytical analysis based on the $k \cdot p$ model, we examine how time-reversal and inversion symmetry operations affect valley-resolved OAM and the associated Berry curvature. In a monolayer QAHI, the $K_+$ valley contributes positive OAM, which under an in-plane electric field moves toward the inner edge. Time-reversal operation reverses both the valley index and the OAM, leading to negative OAM from the $K_-$ valley that accumulates at the outer edge, while inversion symmetry operation alters only the valley index without changing the OAM or its transport direction. When both symmetries operations act simultaneously, the valley index remains unchanged, but the OAM reverses its sign and shifts its accumulation outward. Extending this analysis to bilayer systems with interlayer antiferromagnetic ordering, we consider AA-stacking with mirror plane and AB-stacking with an inversion center. Through spin polarization control, we reveal a spin-driven layer OHE, demonstrating the rich symmetry-governed mechanisms that enable tunable orbital transport in QAHIs.

\section{Discussion}

Starting from the ferromagnetic (FM) monolayer, we construct a general two-band $k \cdot p$ model, which can be written as $H_{eff}=H_0+H_{soc}$. $H_0$ is the term for gapped Direct model, and $H_{soc}$ is the term for spin-orbit coupling (SOC).\cite{RN14, RN575, RN597, RN1028} The detailed Hamiltonian can be written as:
\begin{subequations}\label{eqn-1}
	\begin{gather}
		H_0 = I\epsilon + t (\tau \hat{\sigma}_xk_x+\hat{\sigma}_yk_y)+\dfrac{\Delta}{2} \hat{\sigma}_{z}\\
		H_{soc} = \tau s_{z}
		(\lambda_c \hat{\sigma}_{+}+\lambda_v\hat{\sigma}_{-}) 
	\end{gather}
\end{subequations}
Here $\epsilon$ is a correction energy to set the Fermi level to zero, $\Delta$ is the band gap of the valley, $\tau=\pm1$ is the valley index for $K_+$ and $K_-$ valleys, and  $t$ is the nearest-neighbor intralayer hopping. Moreover, $2\lambda_{c(v)}$ is the band splitting introduced by SOC effect at the conduction band minimum (valance band maximum), and $s_z=\pm1$ represents the spin-up and down channels for z axis. $\hat{\sigma}_i$ is Pauli matrices for pseudospin and  $\hat{\sigma}_{\pm}$ is defined as $\frac{1}{2}(\hat{\sigma}_{0}\pm \hat{\sigma}_{z})$. The Berry curvature(BC)  $\Omega_{n}(\textbf{\textit{k}})$ and orbital Berry curvature (OBC) $\Omega_{n}^{l}(\textbf{\textit{k}})$  at the valley can be calculated by Kubo formula,\cite{RN466}
\begin{equation}\label{eqn-2}
	\begin{aligned}
		\Omega_{n}(\textbf{\textit{k}}) =& 
		-2\hbar^2 \sum_{n\neq n'}
		\frac{{\rm Im}\langle \psi_{n\textbf{\textit{k}}}|\hat{v}_x|\psi_{n'\textbf{\textit{k}}}\rangle	\langle \psi_{n'\textbf{\textit{k}}}|\hat{v}_y| \psi_{n\textbf{\textit{k}}}\rangle}
		{(E_{n'} - E_{n})^2} , \\
		\Omega_{n}^{l}(\textbf{\textit{k}}) =&
		-2\hbar \sum_{n\neq n'}
		\frac{{\rm Im}\langle \psi_{n\textbf{\textit{k}}}|\hat{v}_x|\psi_{n'\textbf{\textit{k}}}\rangle 
			\langle \psi_{n'\textbf{\textit{k}}}|\hat{J}_y| \psi_{n\textbf{\textit{k}}}\rangle}
		{(E_{n'} - E_{n})^2} 
	\end{aligned}
\end{equation} 
Where $\hat{v}_i$ $(i = x, y)$ is the velocity operator along the $k_i$ direction, with $\hat{v}_i=\frac{1}{\hbar} \frac{\partial H}{\partial k_i}$. The orbital current operator $\hat{J}_x$ is defined as $\hat{J}_y = \frac{1}{2}(\hat{v}_y\hat{L}_z+\hat{L}_z\hat{v}_y)$. $\hat{L}_z$ is the $z$ components of the OAM operator.
In our $k \cdot p$ model, the OAM operator of conduction and valance band can be chosen as $\hat{L}_z=\rm{diag}(0,2\hbar\tau)$ for analytical solution of OHE. The detailed BC and OBC for our model is
\begin{equation}\label{eqn-3}
	\Omega_{n}^{\tau}(\textbf{\textit{k}}) = 
	\dfrac{2\tau t^2 \Delta_{\tau,s_z}}{(4t^2k^2
		+\Delta_{\tau,s_z}^2)^{3/2}} = \tau \times \Omega_{n}^{l,\tau}(\textbf{\textit{k}}) 
\end{equation} 
It turns out that, for the out-of-plane spin-polarization, the non-zero BC and OBC is related to the band gap $\Delta_{\tau,s_z}$. Previous work has shown that the OAM distribution at the valleys determines the magnitude of the OHC. If one valley is occupied by states with $\langle \hat{L}_z \rangle = 0\,\hbar$, its OHC vanishes under an applied electric field. In this work, we focus on a QAHI system to study the layer OHE under the operations of time-reversal and inversion symmetries. The band gap in Eq. (3) is set to $\Delta = 0$ eV, indicating the QAHI phase, where the Chern numbers of two valleys are $C_{-1} = C_{+1} = 1/2$. The OBC and BC can be expressed as $\Omega_{n,1}^{\tau}(\textbf{\textit{k}})=\tau \times \Omega_{n,1}^{l,\tau}(\textbf{\textit{k}})=\frac{2s_z t^2 \lambda}{(4t^2k^2+\lambda^2)^{3/2}}$, where $\lambda = \lambda_c-\lambda_v$. At this stage, the OAM distribution and the direction of orbital current under an applied electric field are illustrated in Fig. 1a. Only the $K_+$ valley carries nonzero OAM, which flows laterally under the influence of the BC. As a result, OAM with $\langle \hat{L}_z \rangle > 0\,\hbar$ accumulates along the inner edge of the sample, while no OAM accumulation is observed on the outer edge.

\begin{figure}[!t]
	\centering
	\includegraphics[width=\linewidth]{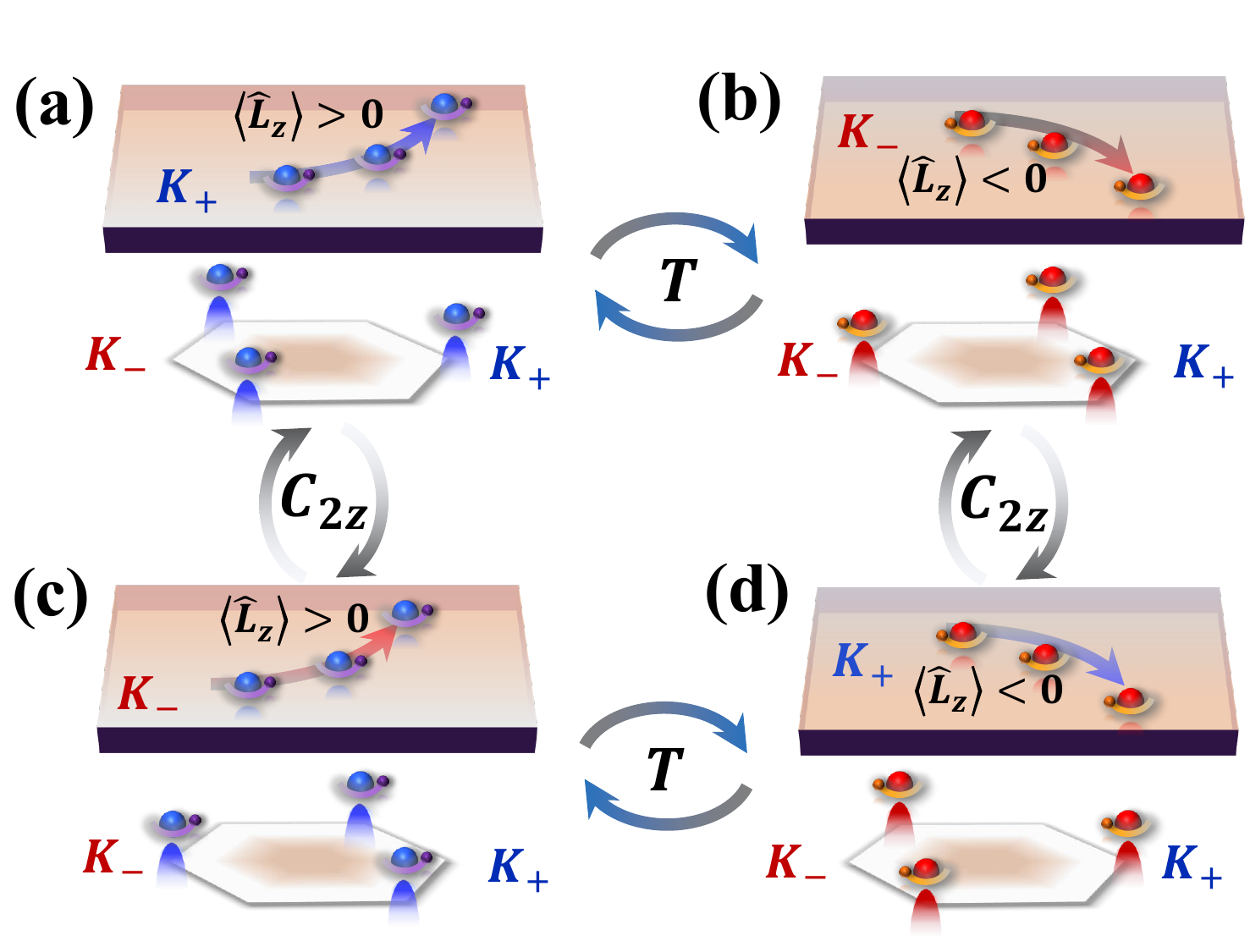}
	\caption{Illustration of OHE in QAHI (a) without operations, (b) with time-reversal operation $\mathcal{T}$, (c) with inversion symmetry operation $\mathcal{I}$ and (d) with the $\mathcal{T}$ and $\mathcal{I}$ operations. The blue and red balls represent the carrier with OAM $\langle \hat{L}_z \rangle > 0 $ and $\langle \hat{L}_z \rangle <0 $, respectively. $K_+$ and $K_-$ are the valley index in the momentum space.}
	\label{fgr1}
\end{figure}
\begin{figure*}[!t]
	\centering
	\includegraphics[width=0.9\linewidth]{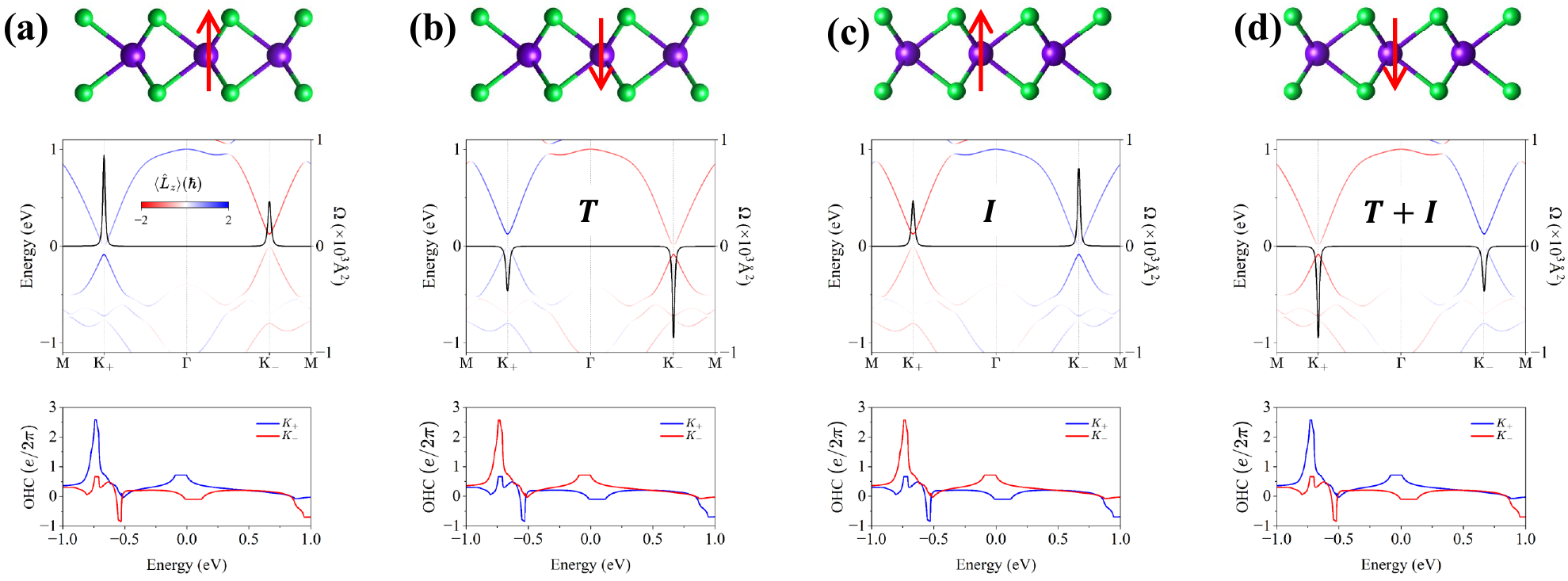}
	\caption{Illustration of OHE in monolayered RuCl$_2$ (a) without operations, (b) with time-reversal operation $\mathcal{T}$, (c) with inversion symmetry operation $\mathcal{I}$ and (d) with the $\mathcal{T}$ and $\mathcal{I}$ operations. The purple and green balls are Ru and Cl atoms, respectively. The red arrow represents the spin polarization.}
	\label{fgr2}
\end{figure*}
RuCl$_2$ crystallizes in space group No. 187 and exhibits two non-degenerate valleys in the first Brillouin zone, denoted as $K_+$ and $K_-$. Previous studies have demonstrated that RuCl$_2$ is a class of Second-order topological insulators exhibiting ferrovalley properties, where the coupling between spin and valley degrees of freedom gives rise to an anomalous valley Hall effect.\cite{RN1214} In addition, in-plane biaxial strain can effectively tune the valley band gap and induce a quantum anomalous Hall state. As shown in Fig. S1, the application of biaxial strain linearly reduces the valley band gap and leads to a band inversion at one of the valleys under 3\% strain, thereby inducing a quantum anomalous Hall effect. The band inversion causes the orbital character of the valley states in RuCl$_2$ to switch from $d_{xy}$ and $d_{x^2-y^2}$ to $d_{z^2}$, leading to a change in the valley OAM expectation value, as shown in Fig. S1(b) and (c). Therefore, as shown in Fig. 2a, the OAM distributions at the $K_+$ and $K_-$ valleys are given by:$\langle \hat{L}_z (K_+) \rangle = +2\hbar, \quad \langle \hat{L}_z (K_-) \rangle = 0\hbar$. 
The magnitude of OHC is closely related to the OAM of the electronic bands. The band inversion further leads to the disappearance of the OHE in the $K_-$ valley. Therefore, the QAHI represents a topological system exhibiting a single-valley OHE, making it an ideal platform for investigating orbital-polarized currents and orbital torques.

Time-reversal operation $\mathcal{T}$ can flip the spin index in the Hamiltonian ($s_z \to -s_z$) and shift the band inversion from the $K_{-}$ valley to the $K_{+}$ valley. Consequently, only the $K_{-}$ valley retains a finite negative OAM ($\langle \hat{L}_z \rangle < 0\,\hbar$), whereas the $K_{+}$ valley no longer hosts any OAM. Under the time-reversal operation $\mathcal{T}$, the BC at the two valleys can be expressed as
\begin{equation}\label{eqn-4}
	\begin{aligned}
		\Omega_{n,2}^{\tau}(\textbf{\textit{k}})= -\dfrac{2s_z t^2 \lambda}{(4t^2k^2+\lambda^2)^{3/2}}, \\
		\langle \hat{L}_z \rangle (K_+) = 0 \, \hbar,\,
		\langle \hat{L}_z \rangle (K_-) = -2\, \hbar, 
	\end{aligned}
\end{equation} 
At this condition, the relationship between the BC and OBC is $\Omega_{n,2}^{l,\tau}(\textbf{\textit{k}})=\tau \times \Omega_{n,2}^{\tau}(\textbf{\textit{k}})$. Therefore, under the applied electric field, the $K_-$ valley carriers with $\langle \hat{L}_z \rangle < 0\,\hbar$ experience a transverse deflection driven by the BC, resulting in the accumulation of OAM along the outer edge of the sample. As shown in comparison with Fig. 2a, reversing the spin polarization changes the valley order of the band inversion, resulting in $\langle \hat{L}_z \rangle = 0\,\hbar$ at the $K_+$ valley and $\langle \hat{L}_z \rangle = -2\,\hbar$ at the $K_-$ valley. The BC, acting as an effective magnetic field in reciprocal space, exerts a Lorentz-like force on carriers perpendicular to the applied electric field. Under the time-reversal operation $\mathcal{T}$, the total OHC remains unchanged, but the valley-resolved contributions are reversed: the $K_-$ valley becomes the primary source of OHC, while the $K_+$ valley no longer contributes. As shown in Fig. 1b, the $\langle \hat{L}_z\rangle< 0\,\hbar$ excited in the $K_-$ valley accumulates at the outer edge of the sample, whereas no OAM accumulation is observed at the inner edge.

Next, we apply the inversion symmetry operation $\mathcal{I}$ to the Hamiltonian in Eq. (1), which corresponds to the transformations $\mathit{k} \rightarrow -\mathit{k}$ and $\tau \rightarrow -\tau$. This operation shifts the band inversion from the $K_-$ valley to the $K_+$ valley. Consequently, the $+2\hbar$ OAM originally associated with the $K_+$ valley in the original system is transferred to the $K_-$ valley.
Under this inversion operation, the BC and the net OAM transform as:
\begin{equation}\label{eqn-5}
	\begin{aligned}
		\Omega_{n,3}^{\tau}(\textbf{\textit{k}}) = \dfrac{2s_z t^2 \lambda}{(4t^2k^2+\lambda^2)^{3/2}}, \\
		\langle \hat{L}_z \rangle (K_+) = 0 \, \hbar,\,
		\langle \hat{L}_z \rangle (K_-) = +2\, \hbar, 
	\end{aligned}
\end{equation} 
Under the $\mathcal{I}$ operation, the calculations of OAM transport are shown in Fig. 1c and Fig. 2c. Notably, although the accumulation still occurs at the same inner edge as in Case 1, the valley origin of the excited OAM shifts from $K_+$ to $K_-$, highlighting the inversion-induced reversal of valley contributions. However, the key distinction lies in the opposite sign of the valley OAM. Consequently, the OAM excited from the $K_-$ valley gives rise to an orbital current that accumulates along the inner edge of the sample with $\langle \hat{L}_z\rangle> 0\,\hbar$, in contrast to the accumulation patterns observed in Figs. 2a and 2b. 

Finally, we consider the last case, in which both $\mathcal{T}$ and $\mathcal{I}$ operations are applied simultaneously, corresponding to the transformations $s \rightarrow -s$, $\mathit{k} \rightarrow -\mathit{k}$, and $\tau \rightarrow -\tau$. Under these combined operations, the BC and OAM in the valleys transform as:
\begin{equation}\label{eqn-6}
	\begin{aligned}
		\Omega_{n,4}^{\tau}(\textbf{\textit{k}})= -\dfrac{2s_z t^2 \lambda}{(4t^2k^2+\lambda^2)^{3/2}}, \\
		\langle \hat{L}_z \rangle (K_+) = -2 \, \hbar,\,
		\langle \hat{L}_z \rangle (K_-) = 0\, \hbar, 
	\end{aligned}
\end{equation} 
As illustrated in Fig. 1d and Fig. 2d, the time-reversal operation $\mathcal{T}$ is applied on top of the inversion operation, resulting in another reversal of valley polarization and shifting the band inversion back to the $K_-$ valley. The electronic structure after applying both $\mathcal{T}$ and $\mathcal{I}$ becomes identical to that of the original configuration in Fig. 2a. However, due to the inversion-induced sign reversal of valley OAM, the $K_+$ valley in Fig. 2d exhibits an OAM contribution opposite to that in Fig. 2a. In this case, although a nonzero OAM is excited at the $K_+$ valley under the applied in-plane electric field, its transport direction is reversed—flowing toward the outer edge and resulting in an accumulation of $\langle \hat{L}_z\rangle< 0\,\hbar$. As a result, under the combined action of $\mathcal{I}$ and $\mathcal{T}$, the system exhibits four fundamentally different OAM transport configurations, highlighting its high degree of tunability and controllability. 

\begin{figure}[!t]
	\centering
	\includegraphics[width=\linewidth]{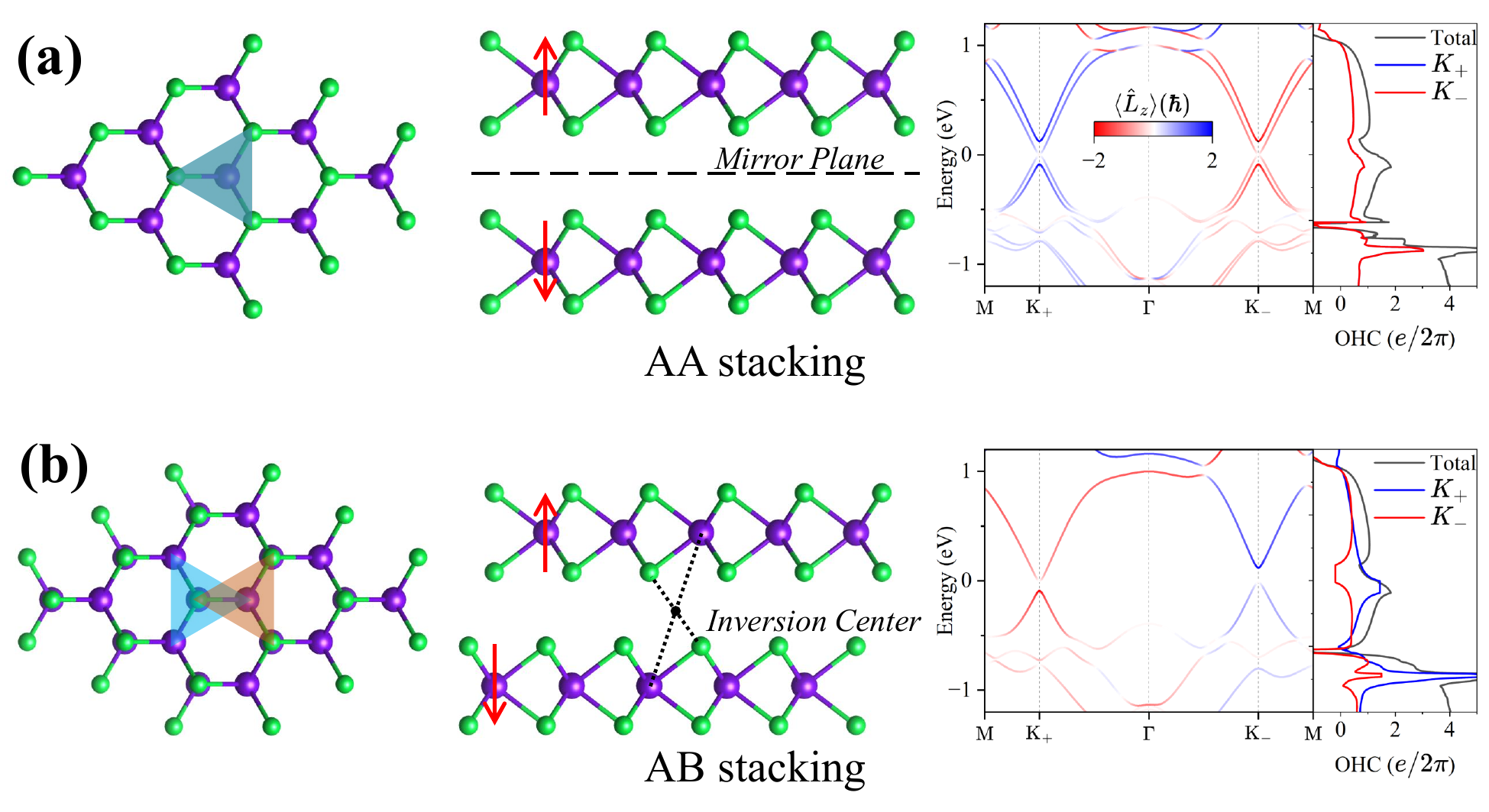}
	\caption{Atomic and electronic structures of RuCl$_2$ bilayer with (a) AA-stacking and (b) AB-stacking. The blue and orange triangles respectively represent the upper layer and the lower layer.}
	\label{fgr3}
\end{figure}

\begin{figure*}[!t]
	\centering
	\includegraphics[width=0.8\linewidth]{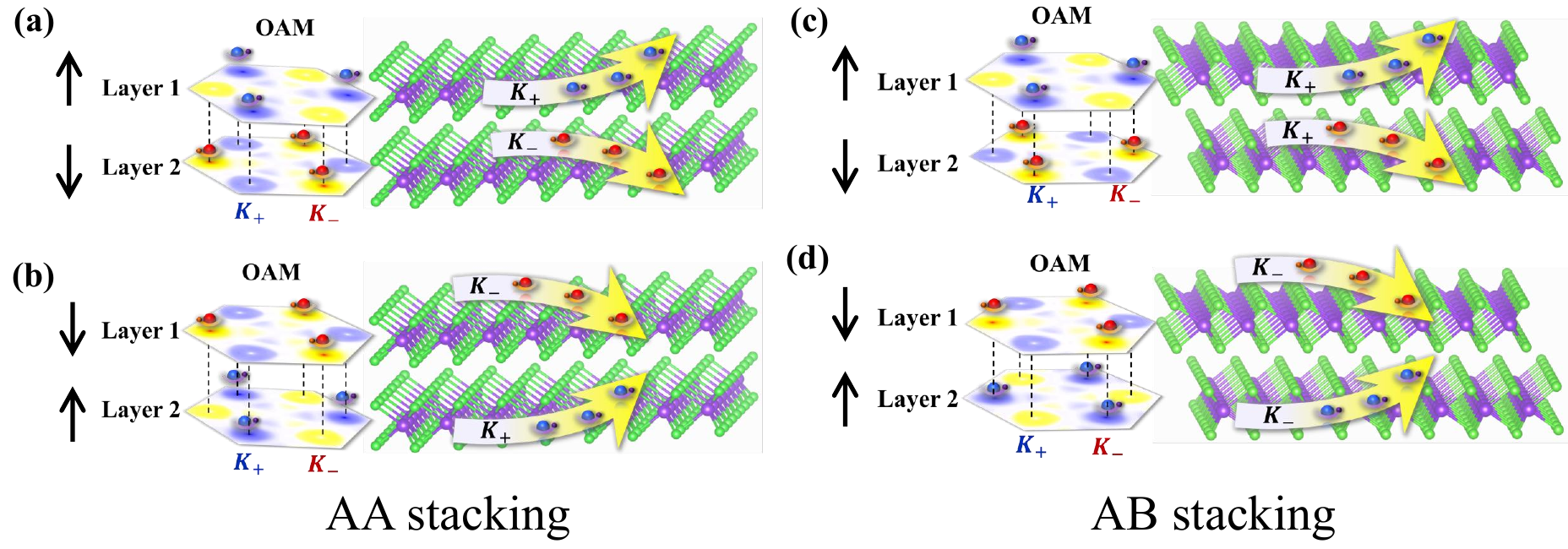}
	\caption{Schematic of layered resolved orbital current in (a,b) AA-stacking and (c,d) AB-stacking. The blue and red balls represent the positive and negative of OAM, respectively, while the yellow arrows indicate the direction of OAM movement. In addition, the black arrows show the spin polarization of each layer.}
	\label{fgr4}
\end{figure*}

However, realizing inversion symmetry $\mathcal{I}$ in a free-standing monolayer is experimentally challenging. To explore the tunability of the OHE in a QAHI, we utilize the stacking flexibility of 2D materials and construct two representative bilayer configurations. The layers are coupled via weak \textit{van der Waals} interactions, which preserve the crystal symmetry and introduce minimal orbital hybridization, thereby allowing symmetry manipulation without significantly altering the underlying electronic structure. In RuCl$_2$ and related TMDCs, bilayers commonly adopt either AA-or AB-stacking, as shown in Figs. 3(a) and 3(b). In the AA configuration, a mirror symmetry plane resides within the \textit{van der Waals} gap, relating the atomic and magnetic structures of the two layers. This symmetry renders the bilayer electronic structure effectively equivalent to a superposition of two monolayers with opposite spin polarizations, analogous to the configurations in Figs. 2(a) and 2(b). Consequently, both the $K_+$ and $K_-$ valleys undergo simultaneous band inversion, suppressing net valley polarization. However, in the absence of inversion symmetry, the valley OAM remains finite and valley-contrasting, yielding equal and additive contributions to the OHC from both valleys. This leads to an enhanced total OHE. Notably, the mirror symmetry maintains independent valley channels for orbital current, while the \textit{van der Waals} gap confines the excited carriers to in-plane motion, limiting interlayer OAM mixing and enabling clean layer-resolved transport. 

As shown in Fig. 4a, the $K_+$ valley in layer 1 generates positive OAM, while the $K_-$ valley in layer 2 generates negative OAM. This leads to the accumulation of positive OAM at the inner edge of layer 1 and negative OAM at the outer edge of layer 2, resulting in a spatial separation of OAM. With the reversal of spin polarization in the bilayer structure, the OAM transport in both layers also changes accordingly. In this case, nonzero OAM is concentrated at the $K_-$ valley and accumulates at the outer edge in layer 1 under a longitudinal electric field. In contrast, in layer 2, the $K_+$ valley contributes nonzero OAM that accumulates at the inner edge. This results in a reversal of the OAM accumulation compared to Fig. 4. The change in spin polarization not only modifies the electronic structure of the system, but also reverses the OAM expectation values and transport directions in the two layers.

The AB-stacking bilayer in Fig. 3(b) possesses a center of inversion symmetry. The valleys from the two layers are fully degenerate and aligned in momentum space, leading to the same valley polarization at both layers. The key difference arises in the layer-resolved OAM: at the $K_+$ valley, both positive and negative OAM contributions coexist, originating from opposite layers. This cancels the net OAM, suppressing global orbital polarization. Nevertheless, the orbital transport remains nontrivial. As illustrated in Figs. 2(a) and 2(d), the opposing OAM signs correspond to opposite transport directions under an applied electric field. These counter-propagating, layer-resolved orbital currents add constructively, leading to an enhanced OHC despite the vanishing net OAM. The orbital polarization is sensitive to symmetry and is typically suppressed by inversion symmetry, whereas orbital transport can remain robust and even become enhanced under the same conditions, as demonstrated in AB-stacking bilayer systems. As a result, in Fig. 4c, this leads to the accumulation of positive OAM at the inner edge of layer 1 and negative OAM at the outer edge of layer 2. With the reversal of spin polarization, the coupling between valleys and orbitals in each layer also switches, consistent with the discussion in Fig. 2. In this case, the dominant orbital current in Fig. 4d shifts from being driven by the $K_+$ valley to the $K_-$ valley. In addition, the sign of the OAM excited at each valley is also reversed. As a result, the orbital current in layer 1 is primarily driven by the $K_-$ valley and accumulates negative OAM at the outer edge, while in layer 2, positive OAM is excited from the $K_-$ valley and accumulates at the inner edge.

Therefore, the orbital transport direction in bilayer antiferromagnetic QAHIs exhibits a pronounced dependence on the spin polarization. The layer-resolved BC can be expressed as $\Omega_{\delta}^{\tau}(\textbf{\textit{k}}) = \delta\frac{2s_0 t^2 \lambda}{(4t^2k^2+\lambda^2)^{3/2}}$ to capture this coupling explicitly, where $\delta=+1 (-1)$ represent the upper (lower) layer, and $s_0$ is the spin polarization of upper layer. Although stacking alters the symmetry of the bilayer system, the interlayer antiferromagnetic ground state enforces opposite spin polarizations between the two layers, leading to layer-contrasting orbital currents and OAM accumulation.
 The presence of an inversion center enforces identical valley polarization in both layers, which in turn causes the OAM in each layer to originate from the same valley. By contrast, when a mirror plane is present, the valley polarization acquires a layer-contrasting character, so that the OAM in the two layers stems from opposite valleys.

\section{Conclusion}\label{secIV}
In summary, we have investigated the orbital transport behavior in bilayer QAHI systems through symmetry analysis based on monolayer. Although stacking modifies the symmetry of the bilayer system, the interlayer antiferromagnetic ground state ensures opposite spin polarizations in the two layers, giving rise to layer-contrasting orbital currents and OAM accumulation. By reversing the spin polarization of the two layers, both the direction and the sign of the OAM in each layer are inverted. In the AA-stacking configuration, mirror symmetry suppresses valley polarization while allowing orbital polarization, giving rise to layer-polarized valley contributions that generate orbital currents in the two layers. In contrast, the AB-stacking configuration, protected by inversion symmetry, preserves intrinsic valley polarization but suppresses orbital polarization, so that the OAM in both layers originates from the same valley. Although the total OHC remains the same in both stacking types, the excitation and transport of OAM are highly sensitive to the system’s symmetry. As a result, by tuning spin polarization, we identify four distinct orbital transport modes, highlighting the crucial role of spin–valley–orbital coupling in determining layer-resolved orbital responses in bilayer QAHIs. This leads to tunable spatial separation of orbital currents between layers, offering a new degree of freedom for designing next-generation orbitaltronic and valleytronic devices.

\begin{acknowledgments}
	This work is supported by National Natural Science Foundation of China (Grant No. 51872145), and Postgraduate Research \& Practice Innovation Program of Jiangsu Province (Grant No. KYCX20\_0748, KYCX19\_0935).
\end{acknowledgments}

\bibliography{apssamp}

\end{document}